\newcommand{\be}{\begin{equation}}
\newcommand{\ee}{\end{equation}}
\newcommand{\bea}{\begin{eqnarray}}
\newcommand{\eea}{\end{eqnarray}}
\newcommand{\lb}{\label}

\documentclass{jltp}

\title{Accelerated Expansion: Theory and Observations}

\author{David Polarski$^*$}

\address{LPM, Univ. Montpellier II, UMR 5825, 34095 Montpellier Cedex 05, 
France\\
$^*$LPMT, UPRES-A 6083, Univ. Tours, F-37200 Tours, France}

\runninghead{David Polarski}{Accelerated Expansion: Theory and Observations}

\begin{document}

\maketitle

\begin{abstract}
The present paradigm in cosmology is the usual Big-Bang Cosmology in which 
two stages of accelerated expansion are incorporated: the inflationary 
phase in the very early universe which produces the classical inhomogeneities 
observed in the universe, and a second stage of acceleration at the present 
time as the latest Supernovae observations seem to imply. Both stages could 
be produced by a scalar field and observations will strongly constrain the 
microscopic lagrangian of any proposed model.

\end{abstract}

\section{INTRODUCTION}
It is interesting that the present paradigm in cosmology contains two phases of accelerated 
expansion of the Universe. The first stage, the Inflationary stage, takes place in the (very) 
early universe and is responsible for the emergence of classical inhomogeneities like temperature 
anisotropies of the Cosmic Microwave Background (CMB) and clustered matter in galaxies.  
The second stage of acceleration takes place today and should have started only recently at 
redshifts $z$ of order 1.
These are combined into the usual Big Bang Cosmology which gains support from two 
observations: the discovery of the CMB and in particular the experimental verification 
with remarkable accuracy of its thermal spectrum; a window of parameters giving the correct 
abundances of light elements in the Big Bang Nucleosynthesis.   

The release of type Ia Supernovae data independently by two groups, the Supernovae Cosmology 
Project and the High-$z$ Survey Project \cite{Perl,Garn}, indicating that our universe is 
presently accelerating, has profound implications on the current paradigm in cosmology.
It was found that, assuming flatness, the best-fit universe with a cosmological 
constant $\Lambda$ is given by $(\Omega_{\Lambda,0},~\Omega_{m,0})=(0.72,~0.28)$. 
If these data are confirmed as could be the case in the near future, they imply a 
radical departure from usual textbooks Friedman-Robertson-Walker cosmology.
Indeed, a perfect isotropic fluid cannot lead to accelerated expansion unless it has a 
sufficiently negative pressure. A pure cosmological constant could of course be responsible 
for this acceleration.
However its amplitude is exceedingly small, about $123$ orders of 
magnitude too small in order to be explained in a natural way (see however \cite{Vol01} for a 
condensed-matter inspired view on this problem). 
This naturalness problem put aside, a cosmological constant is very attractive as it seems 
to make all pieces of our present understanding of structure formation fit into a consistent 
paradigm \cite{BOPS,SS}.
This is why there has been a lot of interest in models where 
the dominant fraction of the present energy density is rather some 
effective (slowly varying) cosmological constant term. 
The prominent candidate in this respect is some (minimally coupled) scalar field 
$\phi$ \cite{RP88,CDS98} often called quintessence, slowly rolling down its potential 
such that it can have a sufficiently negative pressure. 

As for the inflationary phase, its introduction was motivated by basic problems faced by the 
standard picture: the causality problem due to the existence of particle horizons, the flatness 
problem, and finally the origin of the primordial fluctuations from which large scale structures 
in the Universe have emerged through gravitational instability. 
Though an inflationary stage could provide a solution to these problems, a powerful way to put it to 
test is by using observations constraining the primordial fluctuations spectrum. As 
inflationary models make precise predictions, they can be constrained to a large extent and recent 
observations of the CMB anisotropy and of Large Scale Structures, support the inflationary paradigm 
and one can hope that a small class of viable models will meet the observational constraints. 

\section{ACCELERATED STAGE: THEORY}
In cosmology, one assumes that the background space-time has maximally symmetric spatial sections.
The metric is given by
\be
ds^2=dt^2 - a^2(t) \Bigl (\frac{dr^2}{1-kr^2}+ r^2 [d\theta^2 +\sin^2 \theta d\phi^2] \Bigr )~.
\ee 
The values $k=0,1,-1$ correspond respectively to a spatially flat, closed and open universe.
The evolution of the scale factor is governed by the Friedmann equations
\bea
\biggl(\frac{\dot a}{a}\biggr)^2 &=& \sum_{i} \frac{8\pi G}{3} ~\rho_i\lb{FR1} - \frac{k}{a^2}\\
\frac{\ddot a}{a} &=& -\sum_{i} \frac{4\pi G}{3}(\rho_i + 3 p_i)\lb{FR2}
\eea
Here the different matter components labelled $i$, are all isotropic perfect fluids. 
It is convenient to introduce the following dimensionless quantities: the redshift 
$z\equiv \frac{a_0}{a}-1$, the densities relative to the critical density 
$\Omega_{i,0}\equiv \frac{\rho_{i,0}}{\rho_{cr,0}}$ 
where $H=\frac{\dot a}{a},~\rho_{cr,0}\equiv \frac{3H_0^2}{8\pi G}$ and $f_0$ stands for any 
quantity $f$ evaluated today (at time $t_0$), finally $\Omega_{k,0}\equiv -\frac{k}{a_0^2H_0^2}$.

Let us remind first the causality problem in standard Big Bang cosmology. It is easy to show that 
events sufficiently far away in our past light cone were not causally connected at that time. 
Let us take for concreteness two points on the CMB at recombination time, a few hundred thousand 
years after the Big Bang, separated 
by an angle of a few degrees. Looking back, we observe that the CMB has the same temperature in these 
two points: this is a dramatic example of the causality problem because these two points were not 
causally connected at that time. In general, a particle horizon, a distance beyond which no information 
can be received at some given time $t$, exists whenever
\be
\int _0^t \frac{dt'}{a(t')} < \infty ~.
\ee
Using the equations (\ref{FR1},\ref{FR2}) it is easy to see that this will happen in the standard Big Bang 
scenario where the universe is initially radiation dominated with $\rho_{rad}\propto a^{-4}$. 

According to the Friedmann equations, a perfect fluid can induce accelerated expansion provided 
its energy density $\rho$ and its pressure $p$ satisfy the inequality
\be
\rho + 3p < 0~.\lb{in1}
\ee
Clearly, a special type of matter is required for this purpose. An attractive candidate is some 
time dependent scalar field $\phi(t)$ for which
\bea
\rho &=& \frac{1}{2}{\dot \phi}^2 + V(\phi)\\
 p &=& \frac{1}{2}{\dot \phi}^2 - V(\phi)~.
\eea
Hence, under suitable conditions, the inequality (\ref{in1}) can be satisfied in a natural way.

A scalar field, the inflaton, could therefore produce an inflationary stage in the early universe.
This accelerated stage of expansion can arbitrarily enlarge the particle horizon, provided the 
inflationary expansion of the scale factor is arbitrarily large. A value $\frac{a_f}{a_i}\geq 10^{28}$, 
where $a_f$, resp. $a_i$, stands for the scale factor at the end, resp. the beginning, of the 
inflationary stage, is typically required to make all our presently observable universe originate 
from one inflationary patch. In particular, the homogeneity of the CMB, which 
decoupled from the rest of matter after a few hundred thousand years, is now elegantly explained.  
Last, but not least, the quantum fluctuations produced during inflation have their wavelengths 
stretched up to cosmological scales. 
Scales relevant for large scales structure formation and CMB anisotropy,
have their wavelengths stretched far beyond the Hubble radius during inflation and they reenter 
the Hubble radius much later during the radiation and matter dominated eras, depending on the 
scales considered.

Such a stage would dramatically suppress the curvature term $\frac{k}{a^2}$ in the Friedmann equations, 
making it completely negligible at the end of inflation and leading to a spatially flat universe still 
today. 
The spatial geometry can be tested by the CMB anisotopy because, for adiabatic perturbations, the first 
acoustic (Doppler) peak lies roughly at the angular scale of one degree corresponding 
to the angle subtended by the Hubble radius at last scattering. For a flat universe this must 
correspond to $l\approx 200$. The presence of peaks is due to oscillations in the energy density 
(or pressure) of the photon-baryon plasma for scales smaller than the Hubble radius. Hence, at 
recombination, the fluctuation on each such scale is frozen with the corresponding time phase. 
This no longer applies of course if the time phase is stochastic. Generically, independently of the 
precise inflationary model under consideration, the time 
phase of the fluctuations is fixed while (only) the amplitude is stochastic. 
 
Recent CMB anisotropy observations by Boomerang and Maxima of the angular location of 
the first acoustic peak around $l\approx 200$, are in excellent agreement with 
a flat Universe: at two-sigma level one gets $185 \leq l_{peak}\leq 209$.  
Cosmic strings alone (without inflation) lead to a very different spectrum which is excluded by recent 
observations. 

Analogously, a scalar field $\phi$, sometimes coined ``quintessence'', is among the outstanding 
candidates to produce an accelerated stage today. It might well be possible 
in the near future to reconstruct its potential $V$ (in the range corresponding to the ``recent'' 
expansion up to redshifts $z\sim (1-2)$ and its equation of state  
\be
p_{\phi} = w_{\phi}~\rho_{\phi}~,\lb{w}
\ee 
in function of $z$ from luminosity distance measurements, $d_L(z)$
\be 
d_L(z) = (1+z)~H_0^{-1}~|\Omega_k|^{-\frac{1}{2}}~{\cal S}\left( H_0~|\Omega_k|^{\frac{1}{2}}~
\int_0^z \frac{dz}{H(z)}\right)~,\lb{dL}
\ee
where ${\cal S}(u)=\sin u$ for a closed universe, ${\cal S}(u)=\sinh u$ for an open universe 
and ${\cal S}$ is the identity for a flat universe \cite{St98,HT99,SSSS}. This reconstruction can 
actually be extended to non minimally coupled scalar fields in the framework of scalar-tensor 
theories \cite{BEPS00,EP00}.

A crucial point concerns the quantum to classical transition of the fluctuations produced during 
inflation. On those cosmological scales, the squeezing of the fluctuations which are produced 
quantum-mechanically is so huge that it leads to classical stochastic fluctuations \cite{PS1} . 
The nice way to understand this is to use the Wigner function. For the large scale fluctuations, the 
Wigner function will be an elongated ellipse with definite direction in phase-space but negligible width.
Because this width cannot be observed, the fluctuations are indistinguishable from classical stochastic 
fluctuations with stochastic amplitude and fixed time phase. 
This squeezing, which can be computed using some appropriate Bogolyubov transformation, is solely the 
result of the dynamics of the fluctuations on scales bigger than the Hubble radius. 
In the Heisenberg representation it corresponds to the vanishing of the ``decaying'' mode of the 
fluctuations during the evolution outside the Hubble radius.
This result holds for the pure state (isolated system) and can be extended taking into account 
an environment \cite{KPS1}. In this sense, 
inflationary perturbations provide us with a very peculiar system where the squeezing parameter can 
be of order 100 on the largest cosmological scales. 
One should stress again that the amount 
of squeezing makes this system extremely peculiar and certainly no laboratory experiment can hope to 
achieve such huge squeezing. 
Note that a non relativistic free 
particle would experience the same transition to a classical stochastic process \cite{KP}. 
In contrast to a free particle, for cosmological perturbations, one is certainly willing to accept 
the randomness of the fluctuations. Their classicality is expressed in the fact that one can work 
consistently with probabilities along classical trajectories \cite{DP99}. 

\section{OBSERVATIONS}
As said above, SN1a observations have led to the conclusion that our universe is presently 
accelerating and that most of the dark matter is actually smooth and dominates the energy content 
of our universe. 
What is observed is the luminosity distance $d_L(z)$ as a function of redzhift $z$. The luminosity 
distance of an object is defined through
\be
F=\frac{L}{4\pi d_L^2}~,
\ee
where $F$ is the observed flux, and $L$ is the intrinsic luminosity of the object. 
(Actually one works with apparent magnitude $m$, resp. absolute magnitude $M$, corresponding to $F$, resp. 
$L$.)
Assuming that $L$ is known while 
$F$ as well as the redshift $z$ of the object are determined by the observations, one can find the 
luminosity distance $d_L$ as function of $z$. 
The theoretical expression for $d_L(z)$, eq.(\ref{dL}) depends on the type of universe considered.
Assuming that the universe is flat ($k=\Omega_k=0$) and dominated by dust-like matter and a cosmological 
constant, a best-fit universe is obtained for $(\Omega_{\Lambda,0},~\Omega_{m,0})=(0.72,~0.28)$. Such a 
universe is presently accelerating and a pure cosmological constant corresponds to $w=-1$. 
 
Precise observations will enable us to pinpoint the equation of state of this component and its 
possible variation. A pure cosmological constant is not excluded by observations though there is 
considerable uneasiness to cope with such an exceedingly tiny value.
If a scalar field is responsible for the present acceleration, and the relevant part of its potential 
could be recontructed from the observations.
 
As for the fluctuations produced by a specific inflation model, the best constraint will probably 
come from the CMB fluctuations. Great interest is attached to the quantity $C_l$, where 
\be
\langle a^{T*}_{\ell m}a^T_{\ell' m'} \rangle \equiv C^T_{\ell}~\delta_{\ell \ell'}~\delta_{mm'} 
\ee
and the coefficients $a^T_{\ell m}$ are defined through
\be
\frac{\Delta T}{T}= \sum_{\ell=0}^{\infty}\sum_{m=-l}^{m=l}a^T_{\ell m}~Y_{\ell m}~.
\ee  
In an analogous way, the symmetric, trace-free polarization tensor $P_{ab}$ can be 
expanded as follows 
\be
\frac{P_{ab}}{T}=\sum_{\ell=0}^{\infty}\sum_{m=-l}^{m=l}
\Bigl (a^E_{\ell m}~Y_{\ell m,ab}^E +a^B_{\ell m}~Y_{\ell m,ab}^B \Bigr )~,  
\ee
where $Y_{lm}^{E,B}$ are electric and magnetic type tensor spherical 
harmonics, with parity $(-1)^l~{\rm and}~(-1)^{l+1}$ respectively.

For Gaussian fluctuations, which is the case for most inflationary models, all information about the 
CMB anisotropy is contained in the $C_l$'s. They can be computed from first principles for a given 
model. 
At the time of decoupling between radiation and matter the fluctuations are very tiny, of order 
$\sim 10^{-5}$, as we know from the measured CMB temperature anisotropy on large angular scales.
This was actually the major result of COBE DMR, to fix the normalization of the primordial fluctuations. 
Hence the CMB anisotropy can be safely handled using the linearized equations and 
results can be trusted in this regime. The CMB anisotropy measurement will culminate with the Planck (ESA) 
satellite mission to be launched in 2007 which will provide us with a precise measurement of the 
$C_l$'s up to $l\sim 1500$. In addition, it will make a precise mesurement of the CMB polarization 
which will be crucial for breaking to a large extent the degeneracy of the models regarding the CMB 
fluctuations (different parameter combinations can lead to the same $C_l$'s).
The latest CMB observations by Maxima and Boomerang seem to imply that our universe is very nearly 
flat, as testified by the presence of the first acoustic (Doppler) peak at $l\approx 200$, this 
roughly corresponds to the angular scale under which the Hubble radius at decoupling is seen. This 
pronounced peak and its location are the outcome of squeezed adiabatic fluctuations in a flat 
universe and can therefore be seen as giving strong support for the inflationary paradigm. 
Cosmology is certainly having a golden age as more and more observations are carried out with 
increasing accuracy. Future observations will probably reveal us whether our present understanding 
is correct and in case it is, unveil the microscopic lagrangian behind the two accelerated stages 
of expansion. 

\section*{ACKNOWLEDGMENTS}
It is a pleasure to thank the organizers of the ULTI Symposium for the invitation and 
for organizing a most enjoyable meeting.

\end{document}